# LEGAL RESOURCES INFORMATION SYSTEM FOR INFORMATION AGENCIES OF SPECIALIZED LIBRARIES

**Phuc V. Nguyen**, *Ph.D. candidate*
*Lecturer – Hung Vuong University, HCMC, Vietnam*

**INTRODUTION**
In recent years, the rapid development of information technology and communication has a strong impact to industry information - the library. The mission of the industry when in fact the great social place to see the library as knowledge management. Vietnam is in the process of building the rule of law socialist orientation and improves the legal system. So in the current development process, the law library plays an important role in the retention, dissemination and provision of legal information service of legislative, executive and judiciary, particularly especially research, teaching and learning of law school. But the response of the legal information library information agencies remains limited compared to the increasing demand of users.

Lead to the so many different causes, but one very important reason is the body of information, not legal branch libraries to share information with each other. So how to improve service, meet the information needs of legal users believe the problem is the management expertise and concern should be solved. *With significant that the article "Research to share resources for legal information system of library media professional"* with a desire on the basis of research results and propose measures to enhance cooperation to share resources information between the law library in the current context.

The purpose of the research study is based on understanding legal information resources of the library media specialist in Hanoi, as well as the elements necessary and sufficient for the share, works make comments and recommendations to improve the efficiency of resource sharing system for the media agencies specialized law library.

The scope of research is legal information resources of the system of library and information agency specialized in Hanoi such as the Library of Hanoi University of Law, Faculty of Law Resource Center - National University of Hanoi Center Library and Information Science Research of the National Assembly Office, Library of Institute of State and Law Library, Judicial Academy.

The article is focused on the specific content such as demand for exchange of information resources in the system of library and information professionals in law in Hanoi. In depth research that needs to share documents on the system. Like all other types of information resources, information resources and scarce law. Shown in "the ability to provide limited before the actual demand is greater than the ability to provide that information from our users". But the reality is not the media library is also able to meet all information to the user whether the economic capacity allows. Thus, each library, media agencies themselves can not fully meet the needs of users at all times. From this set required to perform active resource sharing of information between library and information agencies. For each agency information, the library has its own characteristics. For example, libraries of Hanoi Law University of users with rich information, with large and mainstream students. With its resources, the level of demand for the system users that have not been satisfied. State Library and Law Institute is an extensive resource intensive legal industry. However, the subjects you read that library service is limited.  There is little to say.  Therefore, the need to share user information inevitably arises. Information users and information needs of library and information agencies specialized legal staff is studying, teaching, students and students of the library.  Type of document readers are most interested in economic law (45%), Criminal Law (70%), Civil Law (40%) ... 100% of readers surveyed answered frequently used documents in Vietnamese. Also, some readers have used the document in English, Chinese (5%), and very small proportion. You read mostly use the form on the spot for reading (95%) and borrowed (100%). Look up information for 65% readers use. Some forms of modern information service users are not interested in mining. Which documents the characteristics of library and information agencies, the existing information resources of the library resources including traditional and electronic materials relatively rich and diverse? Traditional material with hundreds of thousands of books on law, the type of magazine, the system of legal documents... electronic documents to the database directory and online internationally built. About facilities of library and information agencies, the continuous current library is equipped with facilities towards modern uniform. Equipped infrastructure to meet the needs shared computer system connected to internet, modern software reference ... The cooperation of the agencies shared library Hanoi Law University, Resource Center whether the Faculty of Law - Hanoi National University, Centre for Library and Information Science Research of the National Assembly Office, Library of Institute of State and Law - Social

Science Institute of Vietnam, the Library From the Academy new law is just the beginning.

Library of Hanoi University of Law: Activities shared libraries are new have not been formed on a specific policy. Library held workshops with specialized law library; library procures... and has put the issue to share information resources. However, the library has cooperative activities to share the first step in the university library in Ho Chi Minh City Law.

Document sharing activities at the Faculty of Law Resource Center - National University of Hanoi very limited room. The share is mainly conducted with library and information center National University, Hanoi (VNU), Vietnam National library, on a small scale.

Information Center Library and Research of the National Assembly Office shall exchange documents share a number of libraries in the country as the Law Institute of the State Library and Law Library, Justice Department, Library of Hanoi University of Law ... But the exploitation is not shared be conducted regularly and has no official policy in writing.

State Library and Law Institute - Institute of Social Sciences of Vietnam shall co-operation to share the library was carried out with international and domestic for several years, bring more efficiency, increase resources information library. The exchange of library books and magazines to libraries like the Library of Congress, the libraries in France, the library and information agencies in the country.

Library Institute of Justice characterized by its restriction on information resources, facilities,... gallery judicial institutions which have not been shared. Through the survey we find information about resources and activities to share resources information system of library media specialist with strengths in law is the law Vietnam's library a wealth of information resources diversity. Each library has the characteristics and strengths Rien. 2/3 of libraries have been modernized. Our staff of libraries with high professional qualifications. The library has been conducting information sharing resources, sharing activities but also spontaneous and restrictions. In limiting the Law libraries are operating independently, without collaboration sharing of professional activities, sharing resources, and information and library services. The libraries use the management software of different libraries. Libraries are conducting sharing mechanisms do not have a specific policy, limited funds, facilities, poor information infrastructure will be significant barriers to sharing activities take place. We offer a number of measures to improve resource sharing system for the office library and information professionals in law as follows:

**1. Need to build and deploy online union catalogs:**

Building online union catalogs will assist in the interlibrary loans between libraries. Libraries understand each other's information resources, and interlibrary loan transactions will be convenient and fast. See also joint benefit for the parties

**2. For the exchange and cooperation to develop additional sources:**

This form of coordination for resources for any users in the same system as they request. The service users or exchange of documents between the unit members can do directly or indirectly via mail, email or fax, etc. With this method, the number of users as well as financial cycle data of each unit member will be greatly increased, thus the value of information / service efficiency are increased.

**3. Focus on training users:**

Along with surveying the needs of information users in the library and information agencies specialized legal training needs of users, the classes lookup search method is in place. With 50% of readers at the Justice Institute Library, 85% of readers in the Faculty of Law - VNU, 60% of readers in the library of State and Law Institute, 75% of readers in libraries Hanoi Law University needs. Which can be seen, with the deployment of modern services, users can not capture as well as the effective exploitation of such services.

**CONCLUSION**
In summary, tendency to strengthen cooperation to share in the professional associations to enrich the resources are becoming the common trend in all areas of social activity, especially in the field of libraries as information and communication technology Information is increasingly widely used, with the "information explosion", the partnership share it even more necessary than ever. Share information resources as a bridge law will increase the wall strength of the library and information agencies, giving the agency the information library of endless wealth of information, without a single library up can cover it all.